\documentclass[11pt]{article}

\usepackage[T1]{fontenc}
\usepackage[utf8]{inputenc}
\usepackage{lmodern}

\usepackage[margin=1in]{geometry}

\usepackage{amsmath,amssymb}

\usepackage{graphicx}
\usepackage{booktabs}

\usepackage{microtype}

\usepackage[hidelinks]{hyperref}

\usepackage[numbers,sort&compress]{natbib}

\providecommand{\tightlist}{%
  \setlength{\itemsep}{0pt}%
  \setlength{\parskip}{0pt}%
}

\title{OAMAC: Origin-Aware Mandatory Access Control for Practical Post-Compromise Attack Surface Reduction}

\author{Omer Abdelmajeed Idris Mohammed \\ İlhami M. ORAK}

\date{}

\begin{document}

\maketitle

\begin{abstract}
Modern operating systems provide powerful mandatory access control mechanisms, yet they largely reason about \emph{who} executes code rather than \emph{how execution originates}. As a result, processes launched remotely, locally, or by background services are often treated equivalently once privileges are obtained, complicating security reasoning and enabling post-compromise abuse of sensitive system interfaces.

We introduce \textbf{origin-aware mandatory access control (OAMAC)}, a kernel-level enforcement model that treats execution origin---such as physical user presence, remote access, or service execution---as a first-class security attribute. OAMAC mediates access to security-critical subsystems based on execution provenance rather than identity alone, enabling centralized governance over multiple attack surfaces while significantly reducing policy complexity.

We present a deployable prototype implemented entirely using the Linux eBPF LSM framework, requiring no kernel modifications. OAMAC classifies execution origin using kernel-visible metadata, propagates origin across process creation, and enforces origin-aware policies on both sensitive filesystem interfaces and the kernel BPF control plane. Policies are maintained in kernel-resident eBPF maps and can be reconfigured at runtime via a minimal userspace tool.

Our evaluation demonstrates that OAMAC effectively restricts common post-compromise actions available to remote attackers while preserving normal local administration and system stability. We argue that execution origin represents a missing abstraction in contemporary operating system security models, and that elevating it to a first-class concept enables practical attack surface reduction without requiring subsystem-specific expertise or heavyweight security frameworks.
\end{abstract}

\bigskip
\noindent\textbf{Keywords:} Origin-Aware MAC; Kernel Hardening; eBPF LSM; Operating System Security; Provenance-Based Access Control.

\section{\texorpdfstring{\textbf{Introduction}}{Introduction}}\label{introduction}

Remote compromise remains one of the dominant threat vectors in modern
systems. Once an attacker obtains a remote shell---often via SSH---they
typically gain access to the same kernel interfaces available to local
administrators. Interfaces such as \texttt{/sys}, \texttt{/proc/sys},
and kernel debug facilities provide powerful control over system
behavior and are frequently abused for persistence, privilege
escalation, and kernel manipulation.

Traditional Linux security mechanisms---including DAC, capabilities,
SELinux, and AppArmor---primarily focus on \emph{who} a process is, not
\emph{how} or \emph{from where} it was invoked, This leaves execution
origin---such as physical presence versus remote access---outside the
expressiveness of existing access control models. As a result, a root
shell obtained remotely is effectively indistinguishable from one opened
by a physically-present administrator.

We argue that \textbf{execution origin} constitutes a meaningful and
underutilized security signal. In many realistic threat models, physical
presence implies a higher degree of trust than remote access, even when
both operate under the same user identity. This observation motivates
the following research question:

\begin{quote}
\emph{Can Linux enforce mandatory access control based on execution
origin, without kernel patches or heavyweight security frameworks?}
\end{quote}

This paper answers this question affirmatively using eBPF LSM.

Rather than proposing a new access control framework or policy language,
this work identifies \textbf{execution origin} as a missing abstraction
in contemporary operating system security models. Existing mechanisms
are expressive but require administrators to encode provenance
indirectly through labels, domains, or subsystem-specific rules. Our
approach elevates execution origin to a first-class security attribute,
enabling centralized reasoning and enforcement without increasing policy
complexity or altering existing trust assumptions.

\section{\texorpdfstring{\textbf{Threat Model and
Goals}}{Threat Model and Goals}}\label{threat-model-and-goals}

This section clarifies the adversarial assumptions under which OAMAC is
designed and articulates the security objectives it seeks to achieve.
Our goal is not to propose a universal defense against all forms of
system compromise, but to precisely characterize a realistic threat
setting and evaluate how origin-aware mandatory access control can
reduce the attack surface available to adversaries within that setting.
By explicitly stating both the attacker capabilities we consider and
those we exclude, we provide a clear foundation for interpreting the
design, implementation, and evaluation results presented in the
remainder of the paper.

\subsection{\texorpdfstring{\textbf{Threat
Model}}{Threat Model}}\label{threat-model}

\begin{figure}[t]
\centering
\includegraphics[width=0.75\linewidth]{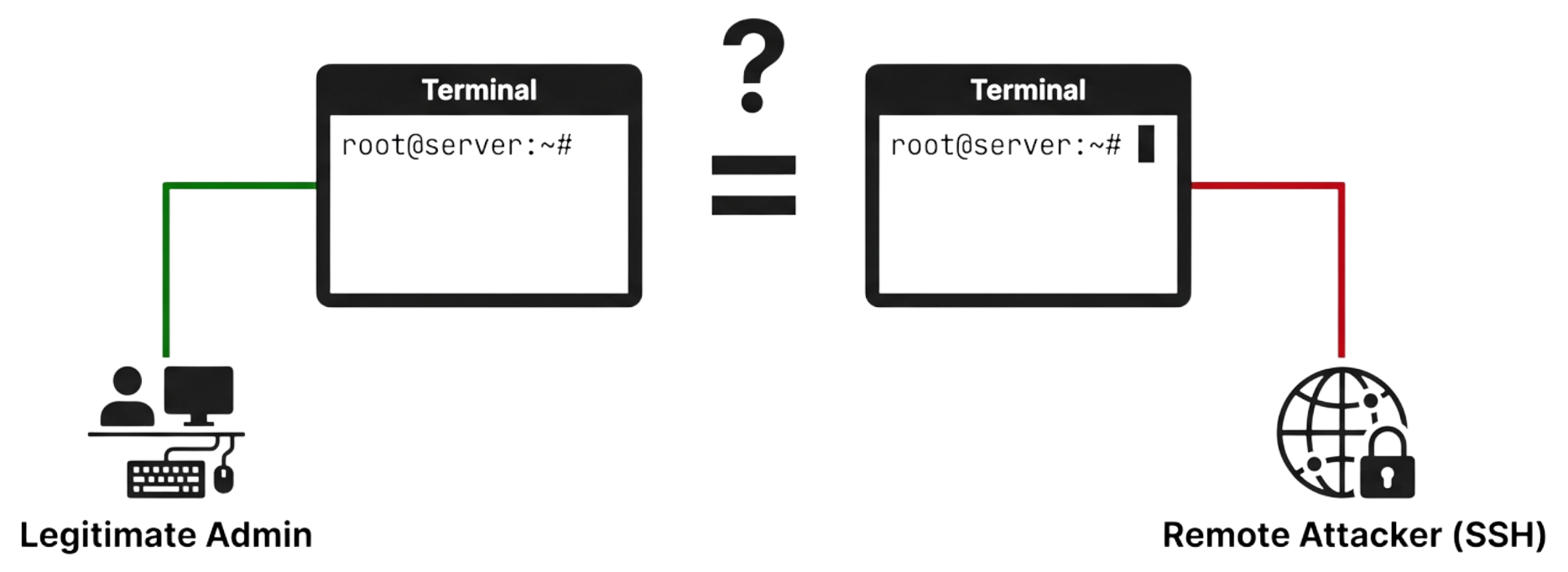}
\caption{Threat model: remote versus physical root.}
\label{fig:threat-model}
\end{figure}

We consider an adversary who gains \textbf{remote access} to the system,
for example through an exposed service or compromised credentials that
allow interactive login such as SSH. The attacker may subsequently
obtain \textbf{root privileges} through misconfiguration, exploitation
of application vulnerabilities, or privilege escalation flaws. However,
we assume that the attacker does \textbf{not} have physical access to
the machine. Figure\textasciitilde{}\ref{fig:threat-model} illustrates
this setting: under traditional Linux access control, a remote root
shell obtained over SSH is effectively indistinguishable from a shell
opened by a physically-present administrator.

This threat model reflects a common and practically relevant
post-compromise scenario in modern server and workstation environments,
where remote access is frequently exposed and privilege escalation
remains a realistic risk. Our focus is therefore on constraining the
actions available to a remote attacker \emph{after} compromise, rather
than preventing compromise itself.

We explicitly exclude attackers with physical access, malicious or
compromised kernels, and hardware-level attacks. As with existing
mandatory access control frameworks, OAMAC assumes a trusted kernel and
enforcement layer. Defending against kernel compromise or physical
tampering is orthogonal to our goals and remains an open problem in
system security.

\subsection{\texorpdfstring{\textbf{Security
Goals}}{Security Goals}}\label{security-goals}

Under this threat model, OAMAC is designed to achieve the following
goals.

First, the system must \textbf{differentiate execution origin inside the
kernel}, distinguishing between processes initiated from physical user
presence, remote access, and background services. This distinction must
be derived from kernel-visible execution provenance rather than
application- or protocol-specific semantics.

Second, OAMAC aims to \textbf{restrict access to security-sensitive
kernel interfaces} for processes originating from remote contexts. The
intent is not to prevent all malicious behavior, but to reduce the set
of high-impact actions available to a remote attacker after privilege
escalation.

At the same time, the system must \textbf{preserve normal local
administration workflows}, including interactive use of \texttt{sudo}
and direct access to system interfaces by physically present users.
Origin-aware enforcement should not introduce friction for legitimate
local management tasks.

From a deployment perspective, OAMAC is designed to \textbf{avoid kernel
patches or custom modules}, relying exclusively on upstream Linux
mechanisms. This enables practical deployment on existing systems and
minimizes maintenance burden.

Operationally, the system must \textbf{support live policy
reconfiguration} without requiring system reboots or process restarts,
and it must remain \textbf{boot-safe}, including during early system
startup when critical services are initialized.

Together, these goals position OAMAC as a practical mechanism for
post-compromise attack surface reduction that complements, rather than
replaces, existing access control systems.

\section{\texorpdfstring{\textbf{Related
Work}}{Related Work}}\label{related-work}

The idea of constraining privileged execution on Linux has been explored
from several angles, including traditional mandatory access control
(MAC) frameworks, eBPF-based security and confinement systems, and
mechanisms that limit post-compromise behavior. This section situates
OAMAC within that landscape. We first review how existing MAC mechanisms
for Linux reason about identity, labels, and objects, and how recent
systematizations still omit execution origin as an explicit
authorization attribute. We then discuss eBPF-based approaches to
security monitoring, kernel hardening, and programmable policy
enforcement, highlighting that these systems mainly treat eBPF itself or
specific subsystems as the protected objects. Finally, we cover work on
post-compromise containment and ``living off the land'' abuse of
legitimate interfaces. Across these domains, we show that prior work
does not elevate execution provenance---such as physical versus remote
sessions or service execution---to a first-class security abstraction, a
gap that OAMAC explicitly targets.

\subsection{\texorpdfstring{\textbf{Mandatory Access Control in
Linux}}{Mandatory Access Control in Linux}}\label{mandatory-access-control-in-linux}

Linux provides several mature Mandatory Access Control (MAC) frameworks,
most notably SELinux, AppArmor, and SMACK \cite{selinux,apparmor,smack}.
These systems associate labels, domains, or profiles with subjects and
objects and enforce policies over subject--object relations. A long line
of work on Linux Security Modules emphasizes this identity- and
label-centric view of authorization \cite{lsm_overview}. More recent
comparative analyses of Linux MAC enforcement mechanisms, including
eBPF-based KRSI extensions, similarly evaluate systems along axes such
as policy expressiveness and performance, but still model decisions as
functions of process identity, labels, capabilities, and object
attributes, without incorporating execution origin or process provenance
as explicit inputs \cite{mac_survey}.

In practice, this means that a shell spawned from a local console and
one obtained via a remote SSH session are treated equivalently unless
administrators manually engineer separate domains or profiles for each
entry point. Our work does not propose another labeling framework;
instead, it introduces \textbf{execution origin} as an additional
authorization dimension that is conceptually orthogonal to these
existing MAC models.

\subsection{\texorpdfstring{\textbf{Limitations of Existing MAC with
Respect to Execution
Origin}}{Limitations of Existing MAC with Respect to Execution Origin}}\label{limitations-of-existing-mac-with-respect-to-execution-origin}

Existing MAC systems fundamentally reason about \emph{what a process is}
(its identity, role, or label) and \emph{which program it executes},
rather than \emph{how the process came into existence}. In SELinux and
AppArmor deployments, once a process has entered a domain or profile,
subsequent access decisions no longer reflect whether it was launched
from a physically-present console, a remote login, a background service,
or a control-plane agent \cite{selinux,apparmor}. While, in principle,
administrators could encode provenance indirectly---e.g., by duplicating
domains for remote and local shells and carefully managing
transitions---such configurations are rare outside bespoke hardening
efforts.

Three factors contribute to this gap.\\
First, \textbf{policy complexity}: encoding origin requires duplicating
rules across applications and entry points, leading to combinatorial
growth.\\
Second, \textbf{fragility}: maintaining origin-specific transitions
across \texttt{fork()}, \texttt{exec()}, and service managers is brittle
under system evolution.\\
Third, \textbf{absence of a first-class signal}: neither classic MAC
frameworks nor eBPF-based MAC extensions expose execution origin or
session provenance as native policy attributes, so provenance can only
be approximated via indirect mechanisms \cite{lsm_overview,mac_survey}.

\begin{figure}[t]
\centering
\includegraphics[width=0.4\linewidth]{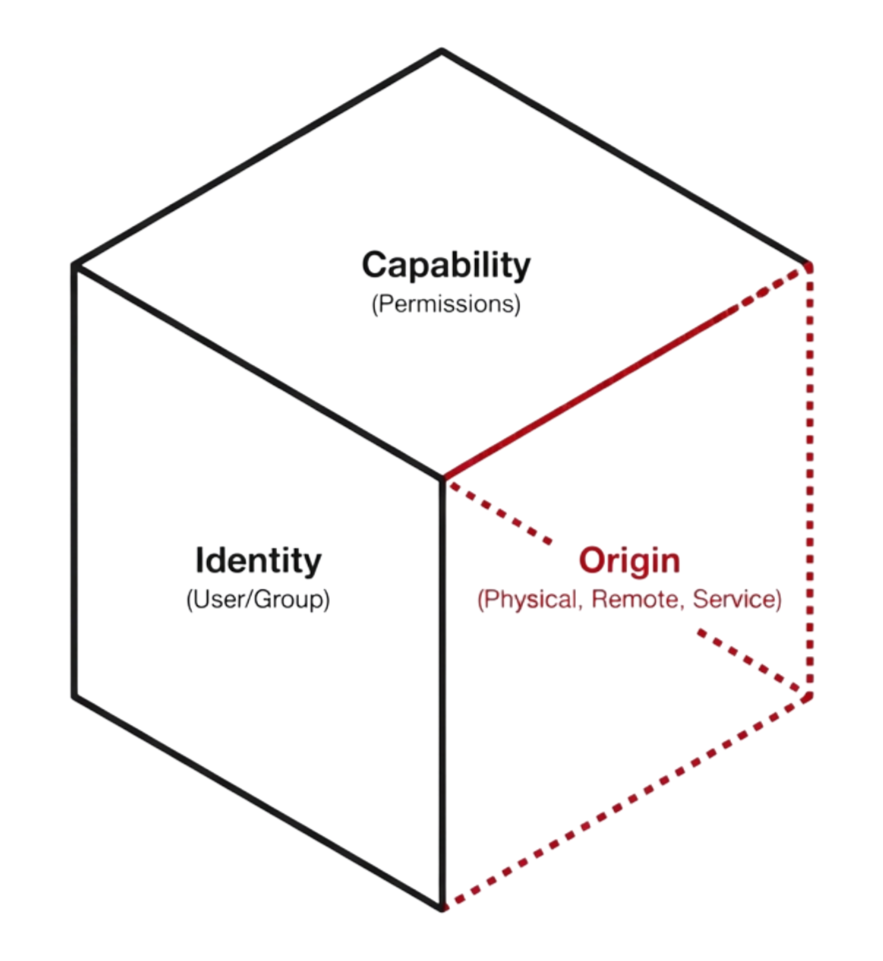}
\caption{Conceptual gap in Linux MAC models.}
\label{fig:oamac-gap}
\end{figure}

Consequently, even recent systematizations of Linux MAC mechanisms do
not identify any design that treats execution origin---such as physical
versus remote sessions or service execution---as a first-class
authorization attribute \cite{mac_survey}. OAMAC explicitly targets this
missing dimension by classifying and propagating origin inside the
kernel and making it directly available to authorization logic.
Figure\textasciitilde{}\ref{fig:oamac-gap} summarizes this conceptual
gap by contrasting identity and capability with origin as an orthogonal
authorization dimension.

\subsection{\texorpdfstring{\textbf{eBPF-Based Security, Hardening, and
Containment}}{eBPF-Based Security, Hardening, and Containment}}\label{ebpf-based-security-hardening-and-containment}

eBPF has become the standard mechanism for safe kernel extensibility and
in-kernel policy enforcement \cite{ebpf_design,bpf_safety}. Production
systems such as Cilium and Tetragon use eBPF for network policy
enforcement and runtime security observability, respectively
\cite{cilium,tetragon}, and prior work has demonstrated eBPF-based
kernel integrity monitoring and syscall tracing for security analysis
\cite{ebpf_security}. These systems largely use eBPF as a flexible
\emph{mechanism} to implement monitoring or hook-specific filtering;
they do not introduce new authorization semantics beyond what existing
MAC models already provide.

A complementary line of work focuses on \textbf{hardening eBPF itself as
an attack surface}. Recent work, including a comprehensive thesis on
eBPF security and defenses, systematizes verifier bugs, helper misuse,
and JIT vulnerabilities and proposes mitigations such as dedicated BPF
LSM programs, runtime integrity checking, and enhanced monitoring
\cite{ebpf_thesis}. Other work, such as SafeBPF, introduces dynamic
in-kernel sandboxing of eBPF programs to guarantee spatial memory safety
even in the presence of verifier bugs, for instance via software fault
isolation or hardware-assisted tagging \cite{safebpf}. These efforts
treat eBPF as the \emph{protected object}: they constrain what eBPF
programs may do and how they access kernel memory. OAMAC instead uses
eBPF LSM hooks merely as an \textbf{implementation vehicle}; the
protected resources are kernel control interfaces and system paths, and
the policy driver is \emph{execution origin}, not the internal behavior
of eBPF programs.

Several systems use eBPF to build \textbf{programmable MAC and
confinement frameworks}. Process-confinement frameworks attach eBPF
programs to multiple hook points and consult per-subsystem maps to
implement flexible, per-process policies over filesystem and networking
operations \cite{ebpf_confinement}. SNAPPY generalizes this idea by
introducing a security namespace that allows containers and other
principals to load their own mandatory eBPF policies into the kernel,
with dynamic helpers and stackable enforcement to support rich, stateful
checks \cite{snappy}. These designs maximize \textbf{policy
expressiveness}: policies are arbitrary eBPF code authored per namespace
or per application. In contrast, OAMAC deliberately avoids a
programmable policy language. It keeps policy \textbf{declarative and
centralized}, exposing only a small, semantically stable set of origins
(e.g., physical, remote, service, bootstrap/unknown) as an additional
input to authorization. Where confinement and SNAPPY ask \emph{``what
checks should we run here?''}, OAMAC asks \emph{``from which origins
should this interface be reachable at all?''}.

Other eBPF-based systems target \textbf{kernel compartmentalization and
stateful behavior enforcement}. Work on on-the-fly kernel
compartmentalization (e.g., O2C) uses eBPF to enforce software fault
isolation and embed machine-learning models inside the kernel to
classify object types during transition, with the goal of containing
vulnerable kernel components \cite{o2c}. Fine-grained runtime
containment agents such as FG-RCA learn benign behavior during an
initial phase, synthesize least-privilege allow-lists, and then use eBPF
LSM hooks to block previously unseen actions at runtime \cite{fg_rca}.
Stateful handling frameworks based on extended finite-state machines
track temporal event sequences (e.g., \texttt{memfd\_create}
\(\rightarrow\) \texttt{execve}) and reactively block or kill processes
once a critical pattern is observed \cite{xfsm}. All of these systems
are \textbf{behavior-centric}: they decide based on \emph{what the code
or process does over time}.

OAMAC is \textbf{origin-centric}. It does not track temporal behavior,
learn allow-lists, or sandbox kernel components. Instead, it classifies
processes by execution provenance (physical console, remote login,
service execution, control-plane helper, bootstrap) using kernel-visible
signals and propagates that origin across \texttt{fork()} and
\texttt{exec()}. Authorization then depends on this \textbf{causal
lineage} rather than on detailed behavioral histories. Systems like
SafeBPF, O2C, FG-RCA, stateful XFSM-based enforcement, and programmable
frameworks such as SNAPPY and eBPF confinement are therefore
\textbf{complementary} to OAMAC: they harden kernel mechanisms or
detect/rule out dangerous behaviors, while OAMAC governs which execution
contexts may reach sensitive interfaces in the first place
\cite{safebpf,o2c,fg_rca,xfsm,snappy,ebpf_confinement}.

Beyond these kernel and eBPF-based frameworks, Optimus focuses on
reducing the attack surface of containerized workloads through
association-based dynamic system call filtering \cite{Optimus2024}.
Optimus tracks program execution phases and enables or disables system
calls based on inferred privilege requirements, thereby shrinking the
temporal window during which sensitive operations are available. In
contrast to OAMAC, which is system-centric and constrains access based
on execution origin, Optimus is program-centric and refines privileges
within a process's lifetime. The two approaches are complementary:
Optimus limits \emph{when} privileged system calls may be exercised,
while OAMAC constrains \emph{from where} such privileged actions are
acceptable in the first place.

\subsection{\texorpdfstring{\textbf{Post-Compromise and
Living-off-the-Land
Attacks}}{Post-Compromise and Living-off-the-Land Attacks}}\label{post-compromise-and-living-off-the-land-attacks}

A substantial body of work documents how attackers abuse legitimate
system interfaces after gaining remote access, often described as
\emph{living off the land} \cite{lolbins}. Common post-compromise
actions include modifying kernel parameters via \texttt{/proc/sys},
writing to \texttt{/sys} to alter kernel or device behavior, and
leveraging BPF interfaces for stealthy persistence or escalation
\cite{bpf_attacks}. Because these interfaces are legitimately accessible
to privileged users, traditional access control and anomaly detection
struggle to distinguish benign administration from malicious
post-compromise behavior.

Existing MAC and eBPF-based monitoring systems can reduce this risk by
tightening policies or detecting suspicious access patterns, but they
generally lack a notion of execution origin
\cite{selinux,apparmor,ebpf_security}. As a result, a remote attacker
who has obtained a privileged shell can often exercise the same
high-impact control paths as a physically-present administrator. By
enforcing access control based on \textbf{execution origin} rather than
identity alone, OAMAC directly targets this class of attacks:
remote-origin processes can be categorically denied access to kernel
control interfaces and BPF control-plane operations, even after
privilege escalation, while physically-present administration remains
unaffected.

\subsection{\texorpdfstring{\textbf{Orthogonal and Complementary
Approaches}}{Orthogonal and Complementary Approaches}}\label{orthogonal-and-complementary-approaches}

Several other mechanisms aim to limit the impact of compromise or
structure security reasoning along different axes, including Linux
capabilities, namespaces and container isolation, and hardware-backed
attestation and trusted execution environments
\cite{linux_capabilities,namespaces,tpm,sgx}. These mechanisms primarily
constrain \emph{what privileges a process holds} or \emph{where code
executes} (e.g., in a container or enclave). Origin-aware MAC is
orthogonal: it constrains \emph{from which execution contexts} certain
powerful interfaces may be used. Combining these
dimensions---identity/labels, containment, hardware roots of trust, and
execution origin---can provide stronger and more intuitive defenses than
any single mechanism alone.

At higher layers, remote desktop systems and UI-level isolation
mechanisms (such as tools that hide sensitive windows from remote
viewers) enforce policies on \textbf{visibility and interaction}, not on
kernel-visible execution provenance. From the kernel's perspective,
processes driven by such tools still execute within a local session and
therefore appear as physical-origin. OAMAC intentionally reasons about
\emph{execution authority} rather than UI semantics, keeping
kernel-level enforcement application-agnostic and allowing UI-layer
controls to coexist as independent defenses.

In summary, prior work has significantly advanced MAC expressiveness,
eBPF safety, behavior-based containment, and stateful detection, but it
has not elevated \textbf{execution origin} to a first-class
authorization attribute. OAMAC fills this conceptual gap by making
origin explicit, persistent, and enforceable inside the kernel, and by
demonstrating how this single abstraction can drive practical attack
surface reduction without new policy languages or kernel modifications.

\section{\texorpdfstring{\textbf{Design
Overview}}{Design Overview}}\label{design-overview}

Modern operating systems already expose rich execution provenance
signals---such as terminal association, session context, and process
ancestry---but these signals remain largely absent from the
authorization model. As a result, existing access control mechanisms
predominantly reason about \emph{who} a process is (identity, labels, or
capabilities), rather than \emph{how} it came to execute. OAMAC
(Origin-Aware Mandatory Access Control) elevates \textbf{execution
origin} to a first-class security attribute by explicitly distinguishing
execution contexts such as physical user presence, remote access, and
background service execution, and using this distinction as a direct
input to authorization decisions. This enables centralized, origin-aware
governance that spans subsystems without entangling authorization logic
with application- or protocol-specific semantics.

At a high level, OAMAC operates in three conceptual stages. First,
execution origin is \textbf{classified} based on observable provenance
signals available at process creation time. These signals reflect the
execution context rather than user identity or network state, enabling
origin to be inferred without protocol-specific assumptions. Second,
once assigned, origin is treated as \textbf{persistent execution
metadata} and is propagated across process creation, ensuring that
descendant processes retain the provenance of their ancestor unless
explicitly reclassified. Third, origin metadata is consulted during
access mediation to enforce \textbf{origin-aware authorization policies}
on security-sensitive system interfaces.

\begin{figure}[t]
\centering
\includegraphics[width=0.75\linewidth]{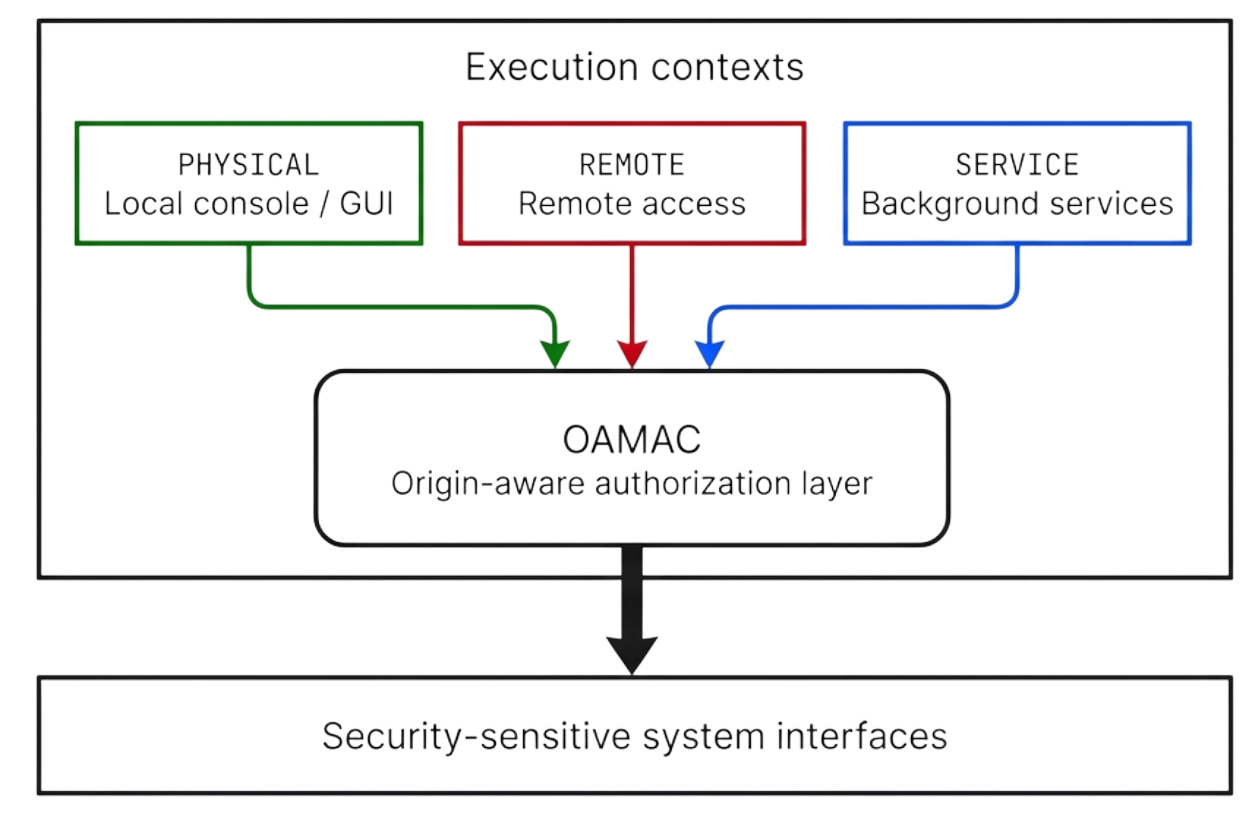}
\caption{Origin-aware MAC architecture.}
\label{fig:arch-overview}
\end{figure}

Figure \ref{fig:arch-overview} illustrates this model at the system
level. Processes enter the operating system through multiple execution
contexts, including interactive local sessions, remote logins, and
service execution. Regardless of entry point, all executions are
associated with a single origin label that remains stable across process
lifetimes. Enforcement is centralized at selected mediation points,
allowing a single origin-aware policy to govern multiple subsystems
consistently.

A key design principle of OAMAC is \textbf{orthogonality}. Origin is not
intended to replace identity- or label-based access control, but rather
to complement existing mechanisms by introducing an additional
authorization dimension. Policies may therefore express constraints such
as \emph{``this operation is permitted only when initiated locally''}
without enumerating users, roles, or domains. This separation allows
OAMAC to reduce policy complexity while remaining compatible with
existing discretionary and mandatory access control frameworks.

Importantly, the design deliberately avoids encoding execution order or
control-flow assumptions between classification, propagation, and
enforcement. These mechanisms operate on distinct system events and are
logically connected only through shared origin state. This abstraction
allows OAMAC to reason about execution provenance independently of
specific operating system subsystems or implementation strategies.

\section{\texorpdfstring{\textbf{Methodology}}{Methodology}}\label{methodology}

This section explains how OAMAC is realized as a concrete Linux
prototype and how its core mechanisms interact in practice. We start by
describing how execution origin is inferred from kernel-visible
provenance signals and propagated across process creation. We then
detail how origin-aware enforcement is implemented at filesystem and BPF
control-plane mediation points, and how these checks are integrated into
the running system without kernel patches or long-lived privileged
daemons. Finally, we outline how policies are configured and applied in
realistic administrative and post-compromise scenarios, providing the
foundation for the evaluation that follows.

\subsection{\texorpdfstring{\textbf{Origin Classification and
Propagation}}{Origin Classification and Propagation}}\label{origin-classification-and-propagation}

\begin{figure}[t]
\centering
\includegraphics[width=0.75\linewidth]{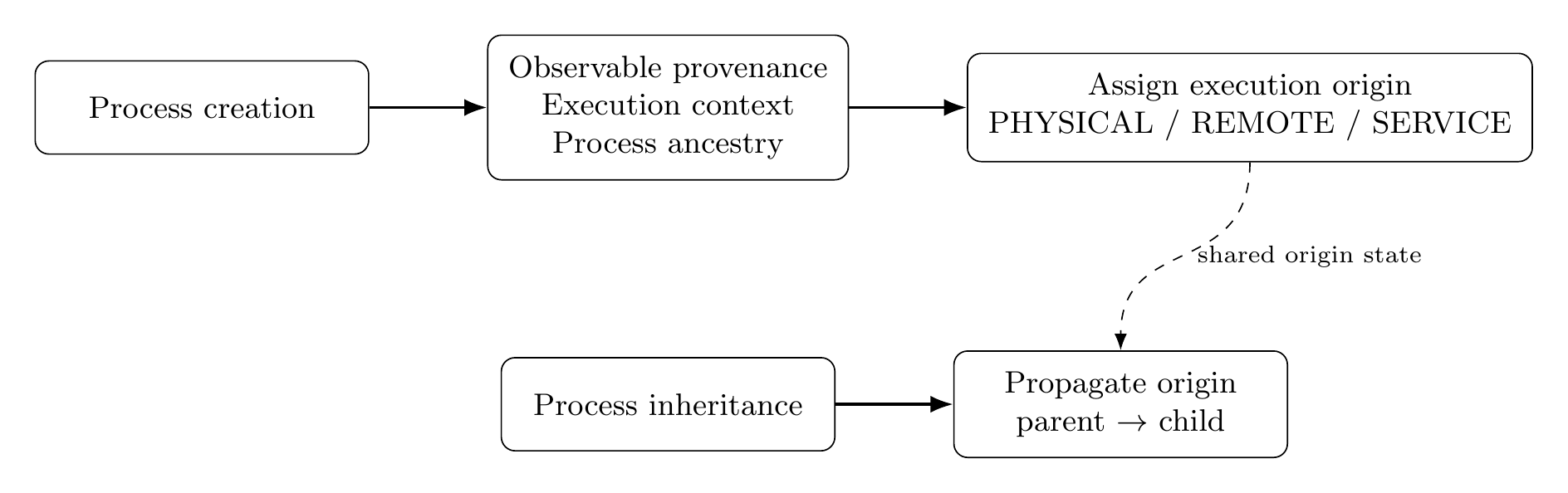}
\caption{Origin classification and propagation.}
\label{fig:origin-classification}
\end{figure}

Figure \ref{fig:origin-classification} details the methodology used to
classify and propagate execution origin. Origin classification occurs at
process execution boundaries, where kernel-visible provenance signals
are examined to infer execution context. These signals include terminal
association and process ancestry, which together allow the system to
distinguish interactive local execution, remote sessions, and
service-initiated processes without relying on network-layer or
application-specific information.

Once classified, origin metadata is stored as shared kernel state
associated with the process identifier. Propagation is handled
independently at process creation events, where child processes inherit
the origin of their parent. The dashed dependency shown in Figure
\ref{fig:origin-classification} denotes a logical relationship through
shared origin state rather than a strict execution sequence, reflecting
the fact that classification and propagation occur on different kernel
paths and may be triggered independently.

This methodology ensures that origin classification is performed once
per execution context and reused consistently across the lifetime of the
process tree. As a result, descendant processes spawned after compromise
or privilege escalation retain the original execution provenance,
preventing origin laundering through process creation.

\subsection{\texorpdfstring{\textbf{Origin-Aware
Enforcement}}{Origin-Aware Enforcement}}\label{origin-aware-enforcement}

\begin{figure}[t]
\centering
\includegraphics[width=0.75\linewidth]{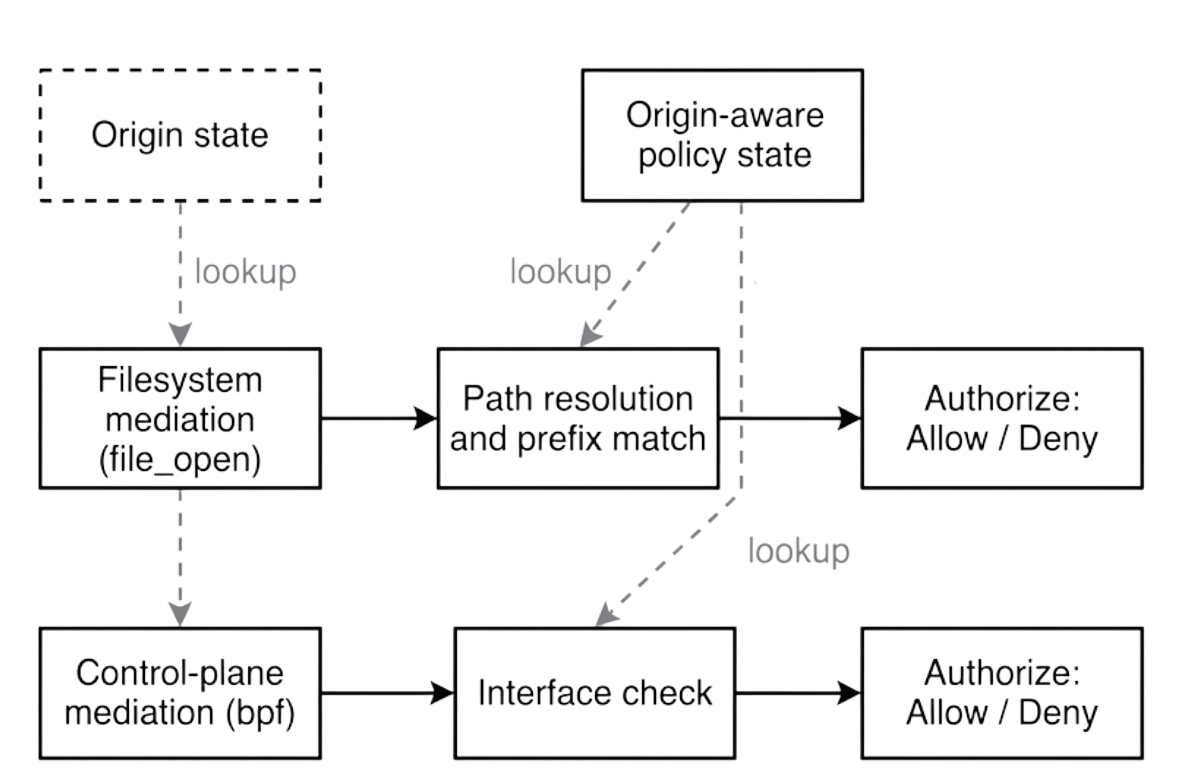}
\caption{Centralized origin-aware enforcement.}
\label{fig:enforcement}
\end{figure}

Figure \ref{fig:enforcement} illustrates the enforcement methodology
applied at selected kernel mediation points. OAMAC adopts a
\textbf{centralized enforcement strategy}, focusing on interfaces that
expose high-impact system control rather than attempting to interpose on
all possible access paths.

In the current prototype, enforcement is applied along two complementary
planes. First, filesystem-mediated interfaces are controlled by
resolving accessed paths and performing prefix-based matching against a
protected set. This allows OAMAC to restrict access to kernel-exposed
control interfaces such as configuration and debugging paths based on
execution origin. Second, kernel control-plane operations are mediated
by enforcing origin-aware authorization on privileged system interfaces,
enabling fine-grained control over operations that modify kernel
behavior or global system state.

In both cases, enforcement decisions are derived from simple
origin-indexed policy lookups. Origin metadata and policy state are
maintained entirely within kernel-resident data structures, allowing
authorization decisions to be made without userspace involvement. Audit
counters are updated on each decision, providing visibility into both
permitted and denied operations while preserving enforcement
determinism.

\subsection{\texorpdfstring{\textbf{System
Integration}}{System Integration}}\label{system-integration}

\begin{figure}[t]
\centering
\includegraphics[width=0.75\linewidth]{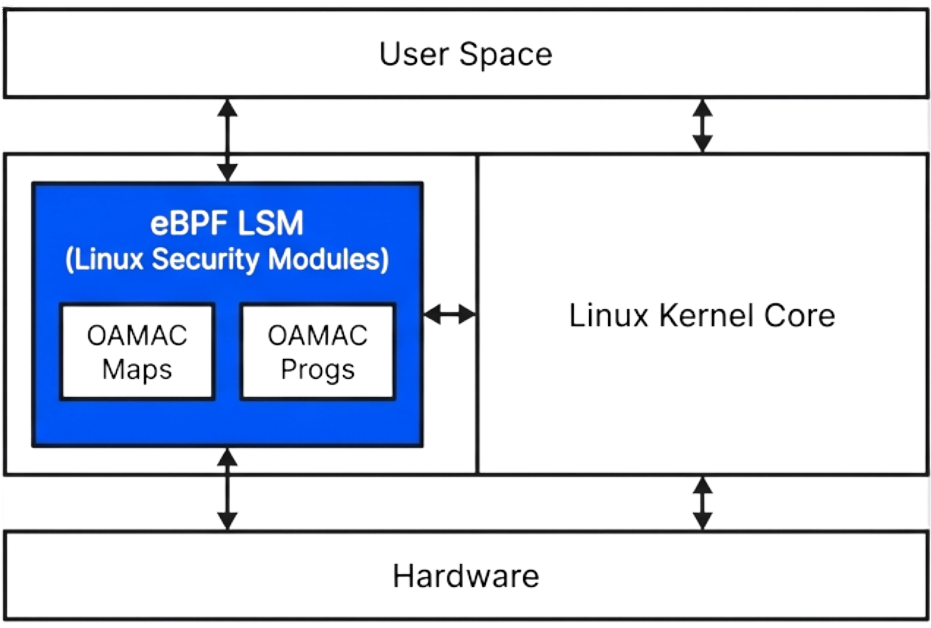}
\caption{OAMAC integration in the Linux kernel.}
\label{fig:oamac-integration}
\end{figure}

As shown in Figure\textasciitilde{}\ref{fig:oamac-integration}, OAMAC is
integrated as an in-kernel enforcement layer that mediates execution
after origin classification but before sensitive operations are
performed. A minimal userspace interface is used solely for policy
configuration and inspection; it does not participate in authorization
decisions. This design ensures that enforcement remains robust against
userspace compromise and avoids introducing long-running privileged
daemons.

By separating classification, propagation, and enforcement into distinct
but coordinated mechanisms, OAMAC achieves origin-aware control over
multiple attack surfaces while maintaining a small and auditable trusted
computing base.

\section{\texorpdfstring{\textbf{Evaluation}}{Evaluation}}\label{evaluation}

This section evaluates whether OAMAC enforces origin-aware mandatory
access control correctly and safely on a real Linux system, without
disrupting normal administration or system operation. Rather than
focusing on synthetic benchmarks or micro-performance measurements, we
adopt a \textbf{behavioral and functional evaluation} aligned with
realistic post-compromise and administration scenarios.

Our evaluation addresses three questions:

\begin{enumerate}
\def\labelenumi{\arabic{enumi}.}
\tightlist
\item
  Does OAMAC correctly distinguish execution origin and enforce policies
  accordingly?
\item
  Does origin-aware enforcement interfere with legitimate local
  administration or system services?
\item
  Does the enforcement mechanism introduce observable overhead or
  operational instability?
\end{enumerate}

\subsection{\texorpdfstring{\textbf{Experimental
Setup}}{Experimental Setup}}\label{experimental-setup}

All experiments were conducted on a Linux virtual machine running a
recent Ubuntu distribution with an upstream kernel supporting eBPF LSM.
The system was configured with:

\begin{itemize}
\tightlist
\item
  eBPF LSM enabled at boot
\item
  \texttt{bpffs} mounted at \texttt{/sys/fs/bpf}
\item
  No kernel patches or custom modules
\item
  No SELinux or AppArmor policy modifications
\end{itemize}

OAMAC was loaded automatically at boot using a \texttt{systemd} oneshot
service. All enforcement decisions were performed entirely in-kernel via
eBPF LSM hooks.

We evaluated three execution origins:

\begin{itemize}
\tightlist
\item
  \textbf{PHYSICAL}: processes launched from a local console or GUI
  terminal
\item
  \textbf{REMOTE}: processes launched via SSH
\item
  \textbf{SERVICE}: system services without an associated TTY
\end{itemize}

All experiments were repeated across multiple reboots to confirm
boot-time stability and reproducibility. In all experiments, unless
otherwise noted, we applied a minimal default policy that denies
high-impact kernel control to non-physical execution while preserving
safe introspection and local administration. Specifically, remote and
service origins were permitted read-only access to kernel BTF metadata
via \texttt{/sys/kernel/btf} to support observability, but remote-origin
access to broader sysfs control surfaces was denied. The OAMAC
configuration directory (\texttt{/etc/oamac}) was protected from both
remote and service origins to prevent runtime policy tampering outside
physically present sessions. At the interface level, BPF control-plane
operations that change kernel behavior---loading programs and creating
or updating maps---were denied for both remote and service origins.
Consequently, post-boot policy changes were performed exclusively from
physically present sessions using the userspace tool, while remote
sessions retained only read-only kernel type introspection.

\subsection{\texorpdfstring{\textbf{Evaluation Scenarios and
Results}}{Evaluation Scenarios and Results}}\label{evaluation-scenarios-and-results}

\subsubsection{\texorpdfstring{\textbf{Remote Post-Compromise
Scenario}}{Remote Post-Compromise Scenario}}\label{remote-post-compromise-scenario}

To simulate a realistic post-compromise setting, we established an SSH
session to the system and executed commands as a privileged user. From
this remote context, we attempted to access security-sensitive kernel
interfaces, including:

\begin{itemize}
\tightlist
\item
  \texttt{/sys}
\item
  \texttt{/sys/kernel}
\item
  selected kernel configuration paths
\end{itemize}

\textbf{Observation.} When origin-aware policies denying remote access
were enabled, all attempts to access protected paths from the SSH
session were denied by the kernel with \texttt{-EPERM}. Denials occurred
consistently at the enforcement boundary, before any subsystem-specific
logic was reached.

\textbf{Result.} OAMAC successfully prevented remote-origin processes
from accessing sensitive kernel interfaces, even when running with root
privileges. This confirms that origin-aware enforcement remains
effective after privilege escalation and aligns with the intended
post-compromise threat model.

\subsubsection{\texorpdfstring{\textbf{Policy Flexibility via
Declarative
Exceptions}}{Policy Flexibility via Declarative Exceptions}}\label{policy-flexibility-via-declarative-exceptions}

We evaluated OAMAC's ability to express flexible policies that broadly
restrict sensitive interfaces while selectively permitting safe,
high-value exceptions for specific origins.

\begin{itemize}
\tightlist
\item
  Setup. We configured broad restrictions (e.g.,
  \texttt{path\ /sys\ deny\ remote} and
  \texttt{path\ /sys\ deny\ service}) and added targeted exceptions
  (e.g., \texttt{path\ /sys/kernel/btf\ allow\ remote} and
  \texttt{path\ /sys/kernel/btf\ allow\ service}) to preserve read-only
  observability.
\item
  Procedure. From a remote SSH session and from a service process,
  attempt:

  \begin{itemize}
  \tightlist
  \item
    Reading \texttt{/sys/kernel/btf/vmlinux}
  \item
    Reading another path under \texttt{/sys} (e.g.,
    \texttt{/sys/kernel/debug})
  \end{itemize}
\item
  Observation. Access to \texttt{/sys/kernel/btf/*} was permitted, while
  other \texttt{/sys/*} paths were denied with \texttt{-EPERM}.
\item
  Result. OAMAC supports least-privilege policy composition:
  administrators can define broad denials and layer narrow, safe
  exceptions to maintain operational needs (e.g., observability) without
  reopening high-impact control surfaces. Implementation detail: because
  policies are evaluated in first-match order, placing exceptions before
  broad rules ensures the intended outcome.
\end{itemize}

\subsubsection{\texorpdfstring{\textbf{Local Administration
Scenario}}{Local Administration Scenario}}\label{local-administration-scenario}

From a local GUI terminal, we executed the same commands under identical
user identities and privilege levels, including the use of
\texttt{sudo}.

\textbf{Observation.} All operations succeeded without restriction.
Origin classification remained stable across privilege escalation, and
no unexpected denials or behavioral changes were observed during
interactive use.

\textbf{Result.} OAMAC did not interfere with legitimate local
administration. This confirms that origin-aware policies can restrict
remote attack surfaces while preserving normal local workflows,
including privileged operations.

\subsubsection{\texorpdfstring{\textbf{Service Stability and Boot
Safety}}{Service Stability and Boot Safety}}\label{service-stability-and-boot-safety}

We evaluated whether origin-aware enforcement affected system services
or boot-time execution.

\textbf{Observation.} The system booted successfully with OAMAC enabled.
Standard system services continued to operate normally, and no service
failures or startup regressions were observed. Enforcement did not
affect non-interactive workloads.

\textbf{Result.} OAMAC is boot-safe and does not disrupt service
execution, satisfying a core deployment requirement for mandatory access
control mechanisms.

\subsubsection{\texorpdfstring{\textbf{Loader Service Restart Under
Lockdown (SERVICE-Origin
Enforcement)}}{Loader Service Restart Under Lockdown (SERVICE-Origin Enforcement)}}\label{loader-service-restart-under-lockdown-service-origin-enforcement}

We evaluated restarting the systemd loader after applying strict
interface denies to SERVICE origin.

\begin{itemize}
\tightlist
\item
  Setup. The default policy included
  \texttt{iface\ bpf-prog-load\ deny\ service},
  \texttt{iface\ bpf-map-create\ deny\ service}, and
  \texttt{iface\ bpf-map-update\ deny\ service}. The loader runs without
  a controlling TTY and is therefore classified as SERVICE.
\item
  Observation. Restarting the loader with these denies in place caused
  the unit to fail, as expected: the loader's BPF operations were
  blocked at the lsm/bpf enforcement point.
\item
  Result. This confirms correct classification (SERVICE) and enforcement
  for control-plane operations originating from services. Operationally,
  maintenance can be performed from a PHYSICAL session by temporarily
  removing SERVICE BPF denies via \texttt{oamacctl}, restarting the
  loader, and then re-applying the denies.
\end{itemize}

\subsubsection{\texorpdfstring{\textbf{Policy Reconfiguration and
Control-Plane
Protection}}{Policy Reconfiguration and Control-Plane Protection}}\label{policy-reconfiguration-and-control-plane-protection}

We evaluated dynamic policy updates using the \texttt{oamacctl} control
tool.

The evaluation included:

\begin{enumerate}
\def\labelenumi{\arabic{enumi}.}
\tightlist
\item
  Enabling origin-based restrictions for both filesystem paths and BPF
  control-plane operations
\item
  Verifying immediate enforcement without restarting processes
\item
  Disabling restrictions and confirming access restoration
\end{enumerate}

We additionally installed interface policies that disallowed BPF program
loading and map updates from remote origins.

\textbf{Observation.} Policy changes took effect immediately.
Remote-origin processes were prevented from loading new BPF programs or
modifying policy maps, while physically-present sessions retained full
control.

\textbf{Result.} OAMAC supports safe live reconfiguration and prevents
remote attackers from weakening enforcement post-compromise, even if
they temporarily obtain root privileges.

\subsection{\texorpdfstring{\textbf{Correctness
Validation}}{Correctness Validation}}\label{correctness-validation}

Across all scenarios, correctness was evaluated using the following
criteria:

\begin{itemize}
\tightlist
\item
  \textbf{Origin classification accuracy}: processes were consistently
  labeled as PHYSICAL, REMOTE, or SERVICE according to their execution
  context
\item
  \textbf{Enforcement correctness}: access decisions aligned with
  configured origin-aware policies
\item
  \textbf{Non-interference}: allowed workflows continued to function
  normally
\item
  \textbf{Boot safety}: the system remained bootable with enforcement
  enabled
\end{itemize}

Kernel-resident audit counters were used to confirm that enforcement
hooks were triggered as expected and that both permitted and denied
operations were correctly recorded.

\subsection{\texorpdfstring{\textbf{Overhead and Performance
Considerations}}{Overhead and Performance Considerations}}\label{overhead-and-performance-considerations}

While performance benchmarking was not a primary goal, we qualitatively
assessed overhead during interactive use.

\textbf{Observation.} No noticeable performance degradation was
observed. Enforcement adds only constant-time checks in LSM hooks and
introduces no persistent userspace daemons.

\textbf{Result.} OAMAC imposes minimal overhead and does not affect
system responsiveness under typical workloads.

\subsection{\texorpdfstring{\textbf{Reproducibility}}{Reproducibility}}\label{reproducibility}

All components of OAMAC---including kernel-resident eBPF LSM programs,
userspace control utilities, and boot-time integration scripts---are
deterministic and reproducible. The complete prototype implementation
used in this evaluation is released as open-source software and publicly
available online \cite{oamac2026}.

The evaluation can be replicated on any Linux system with upstream eBPF
LSM support enabled, without kernel modification, custom modules, or
external dependencies beyond standard eBPF tooling. All policy
configurations, enforcement logic, and control-plane operations
described in this paper correspond directly to the released
implementation.

\section{\texorpdfstring{\textbf{Discussion}}{Discussion}}\label{discussion}

This section discusses the broader implications of origin-aware
mandatory access control beyond the concrete prototype evaluated in this
work. We examine how OAMAC relates to existing access control
mechanisms, explore design trade-offs and deployment considerations, and
consider how the underlying abstraction could be integrated more deeply
into operating system kernels and extended to other environments. These
discussions contextualize the design choices made in this paper and
clarify both the strengths and the boundaries of origin-aware
enforcement as a system security primitive.

\subsection{\texorpdfstring{\textbf{Kernel-Integrated Origin as a Design
Extension}}{Kernel-Integrated Origin as a Design Extension}}\label{kernel-integrated-origin-as-a-design-extension}

The current OAMAC prototype is implemented using upstream Linux features
without kernel modification, relying on eBPF LSM hooks and
kernel-visible execution provenance to classify processes at runtime.
This design choice prioritizes deployability and allows OAMAC to be
evaluated on unmodified systems. However, it also introduces an inherent
limitation: processes that exist before OAMAC is initialized---such as
early system tasks and boot-time services---cannot be retroactively
classified and are conservatively assigned an \texttt{UNKNOWN} origin.

To preserve boot safety and avoid disrupting existing system behavior,
origin-aware policies in the current design are not applied to
\texttt{UNKNOWN} origins by default. While this is a safe and practical
choice, it conflates two distinct cases: early bootstrap execution that
must remain unrestricted, and later execution contexts whose provenance
is genuinely ambiguous. This overloading limits the ability to apply
restrictive policies to unknown or suspicious execution contexts without
risking interference with core system processes.

A native kernel integration of OAMAC would allow this limitation to be
addressed cleanly. In such a design, execution origin would be
represented as a first-class field in the kernel's process metadata
(e.g., within \texttt{task\_struct}) and assigned at task creation time.
The kernel could explicitly distinguish early bootstrap execution using
a dedicated origin such as \texttt{BOOTSTRAP}, assigned exclusively
during system initialization. This origin would be exempt from
origin-aware enforcement, preserving boot safety by construction.

Once the system reaches a stable operational state and origin-aware
policy is active, no new tasks would be permitted to enter the
\texttt{BOOTSTRAP} origin. Processes created thereafter would be
classified as \texttt{PHYSICAL}, \texttt{REMOTE}, \texttt{SERVICE}, or
\texttt{UNKNOWN} based on kernel-visible provenance available.
Importantly, this separation allows \texttt{UNKNOWN} origins to be
treated as policy-controlled and potentially restricted, without
affecting early system tasks or undermining system startup semantics.

Beyond resolving the bootstrap classification ambiguity, a
kernel-integrated design offers additional benefits. Origin propagation
becomes invariant and race-free, authorization checks no longer depend
on auxiliary data structures, and origin can be exposed uniformly to all
kernel authorization mechanisms. More broadly, elevating execution
origin to a first-class kernel attribute clarifies its semantics and
enables principled reasoning about origin-aware policies across
subsystems.

While this work does not implement such kernel modifications, we argue
that introducing execution origin into the kernel's core process
representation is a natural and valuable extension of existing security
models. The proposed design preserves the deployability advantages of
the current prototype while outlining a clear path toward stronger
semantics, finer-grained policy control, and cleaner integration with
existing mandatory access control frameworks.

\subsection{\texorpdfstring{\textbf{Relation to Existing MAC
Systems}}{Relation to Existing MAC Systems}}\label{relation-to-existing-mac-systems}

OAMAC differs from traditional mandatory access control systems such as
SELinux and AppArmor in both its scope and its abstraction level.
Existing MAC frameworks primarily regulate access by associating labels,
domains, or profiles with subjects and objects, and by expressing
policies that reason about \emph{who} a process is or \emph{which
program} it executes. While these mechanisms are powerful and
expressive, they require explicit labeling and often entail complex,
subsystem-specific policy engineering.

In contrast, OAMAC does not introduce new labels on files or processes,
nor does it define a new policy language. Instead, it introduces
\textbf{execution origin} as an additional authorization dimension that
is orthogonal to identity- and label-based controls. Origin-aware
policies reason about \emph{how execution was initiated}---for example,
whether a process originates from physical user presence, remote access,
or background service execution---rather than about application identity
alone.

Importantly, OAMAC is not intended to replace existing MAC frameworks.
Its design is explicitly orthogonal: origin-aware enforcement can
coexist with SELinux or AppArmor and can be applied independently of
their policies. Because OAMAC does not require relabeling objects or
restructuring existing policies, it can be layered alongside established
MAC systems to constrain execution context without requiring policy
refactoring or kernel modification.

\subsection{\texorpdfstring{\textbf{Input Devices and Physical
Presence}}{Input Devices and Physical Presence}}\label{input-devices-and-physical-presence}

Our current origin classification does not directly distinguish between
different types of physical input devices (e.g., built-in keyboard, USB
keyboard, virtual keyboard). Instead, it relies on \emph{kernel-visible
execution context}---specifically TTY association and process ancestry.

From a security perspective, this is a deliberate design choice. The
Linux kernel does not reliably expose the provenance of input events
(physical vs virtual) to security hooks, and attempting to infer trust
from input hardware would introduce fragile and platform-specific
assumptions.

Importantly, our threat model treats \emph{local kernel execution
context} as the trust boundary, not the input device itself. Whether
commands are entered via a physical keyboard, USB device, or
accessibility interface, the resulting process is still mediated by the
same local TTY and process hierarchy.

We argue that origin-aware MAC should reason about \emph{execution
provenance}, not human interaction modality. This keeps the design
robust, auditable, and compatible with existing kernel abstractions.

\subsection{\texorpdfstring{\textbf{Generalization to Other Operating
Systems}}{Generalization to Other Operating Systems}}\label{generalization-to-other-operating-systems}

Although the current implementation of OAMAC targets Linux, the
underlying concept of origin-aware access control is not tied to a
specific operating system. Most modern operating systems already
maintain implicit execution provenance information. For example, Windows
distinguishes between interactive logons, service execution, and remote
sessions; macOS tracks console sessions and remote login contexts; and
mobile operating systems differentiate between foreground user actions
and background services. These distinctions are routinely used for
session management and auditing, but they are rarely exposed as
first-class inputs to kernel-level authorization decisions.

An origin-aware design could be generalized across platforms by making
execution origin an explicit and persistent attribute of processes. This
would involve defining a small, fixed set of origin categories---such as
bootstrap, physical, remote, service, and unknown---propagating origin
metadata automatically with process creation, and allowing security
policies to depend on origin in addition to traditional identity-based
attributes. Importantly, such a design does not require rethinking
existing access control models; rather, it augments them with an
additional, orthogonal authorization dimension.

We argue that the absence of execution origin from the authorization
layer represents a structural gap in contemporary operating system
security models. Elevating origin to a first-class security attribute
enables more expressive and intuitive policies and provides a unifying
abstraction for governing execution context across subsystems and
platforms.

\subsection{\texorpdfstring{\textbf{User-Centric Governance and
Centralized Security
Reasoning}}{User-Centric Governance and Centralized Security Reasoning}}\label{user-centric-governance-and-centralized-security-reasoning}

A recurring challenge in system security is that effective protection
often requires deep, subsystem-specific expertise. Securing filesystems,
kernel interfaces, device nodes, networking stacks, and runtime services
typically demands detailed knowledge of each component's threat model
and configuration semantics. This complexity exceeds what is
realistically expected from normal users, and even from many advanced
administrators.

Our design intentionally shifts security reasoning away from individual
subsystems and toward a \emph{user-centric governance layer} based on
execution origin. Instead of requiring users to understand the security
implications of each protected interface, the system allows users to
express a simpler and more intuitive intent: \textbf{which subsystems
should be accessible from which execution origins}.

For many users, especially in server and workstation environments, this
intent is straightforward. Users often do not require remote access to
sensitive kernel interfaces, device control paths, or system
configuration subsystems. By allowing users to declare that such
subsystems should be accessible only from physically present sessions,
the system can significantly restrict a large class of potential attack
surfaces with minimal policy complexity.

This origin-centric abstraction provides several advantages:

\begin{enumerate}
\def\labelenumi{\arabic{enumi}.}
\item
  \textbf{Reduced Cognitive Load}\\
  Users no longer need to reason about the security properties of each
  individual subsystem. Instead, they reason about trust in execution
  context (local vs remote vs service), which aligns more closely with
  human intuition and operational practice.
\item
  \textbf{Broad Attack Surface Reduction}\\
  Because many high-impact post-compromise attacks rely on access to
  kernel and system interfaces reachable via \texttt{/sys},
  \texttt{/proc/sys}, device nodes, or similar control planes,
  restricting these paths by origin significantly reduces the feasible
  actions available to a remote attacker.
\item
  \textbf{Centralized Monitoring Point}\\
  All protected subsystems are mediated by a \textbf{logically
  centralized} origin-aware enforcement layer. As a result, monitoring
  the health and behavior of this layer provides a centralized
  visibility point. Any attempt to disable, bypass, or modify the
  origin-based policy becomes a strong indicator of an ongoing attack
  targeting protected system internals.
\item
  \textbf{Compositional Safety Guarantees}\\
  While no single mechanism can guarantee absolute system security,
  origin-aware enforcement enables a strong compositional property:
  \textbf{as long as the origin layer remains correct and uncompromised,
  all subsystems protected by it inherit its security guarantees with
  respect to execution origin}. This allows users and administrators to
  reason about system safety at a higher level of abstraction.
\item
  \textbf{Practical Security for Real Users}\\
  The design acknowledges that security mechanisms are only effective if
  they are usable and understandable. By providing a simple, centralized
  control plane, the system enables users to apply meaningful security
  restrictions without requiring specialized domain knowledge.
\end{enumerate}

We argue that this shift---from subsystem-specific hardening to
origin-based governance---represents a practical and scalable approach
to reducing attack surfaces in modern operating systems.

\subsection{\texorpdfstring{\textbf{Distinguishing Cloud Control-Plane
Access}}{Distinguishing Cloud Control-Plane Access}}\label{distinguishing-cloud-control-plane-access}

In cloud environments, not all remote access is equal. Cloud providers
often offer management access through authenticated control-plane
interfaces, which differ fundamentally from direct SSH access initiated
by users.

Our design accommodates this distinction by treating control-plane
access as a separate execution origin. Rather than attempting to infer
trust from network-level properties, the system relies on explicit
signals provided by trusted components, such as PAM modules, system
services, or provider-specific agents.

For example, a cloud provider may authenticate a user through its
control panel and launch a session that explicitly marks the resulting
process as originating from the control plane. This marker can be
communicated to the kernel via trusted system components, such as
authenticated services or provider-specific agents, allowing
origin-aware enforcement to distinguish control-plane sessions from
arbitrary remote logins.

This approach enables cloud operators to grant limited, auditable access
to sensitive subsystems for management purposes while maintaining strict
restrictions on general remote access. Importantly, it preserves the
centralized governance model: users and providers reason about trust at
the origin level, not at the level of individual subsystems.

\subsection{\texorpdfstring{\textbf{Handling Non-SSH Remote
Access}}{Handling Non-SSH Remote Access}}\label{handling-non-ssh-remote-access}

Our design does not rely on SSH-specific semantics. Instead, it
classifies execution origin based on kernel-visible provenance,
including terminal association and process ancestry.

As a result, a wide range of remote access mechanisms---including
reverse shells, web-based management interfaces, container exec
operations, and agent-mediated access---are naturally mapped to
non-physical origins. In many cases, such mechanisms are classified more
restrictively than interactive SSH sessions, as they lack a physical
terminal context.

This abstraction avoids protocol-specific assumptions and ensures that
origin-aware enforcement remains effective even as remote access
techniques evolve.

\section{\texorpdfstring{\textbf{Limitations}}{Limitations}}\label{limitations}

While origin-aware access control provides a practical and effective
mechanism for reducing attack surfaces, it is subject to several
limitations that are important to acknowledge.

\subsection{\texorpdfstring{\textbf{Trust in the Kernel and Enforcement
Layer}}{Trust in the Kernel and Enforcement Layer}}\label{trust-in-the-kernel-and-enforcement-layer}

Our design assumes the integrity of the kernel and the correctness of
the origin-aware enforcement mechanism. As with all mandatory access
control systems, a successful kernel-level exploit, rootkit, or
arbitrary kernel memory corruption could bypass or disable the
enforcement logic.

This limitation is not unique to our approach and is shared by existing
MAC frameworks such as SELinux and AppArmor. Defending against kernel
compromise is orthogonal to our goals and remains an open challenge in
system security.

\subsection{\texorpdfstring{\textbf{Granularity of Origin
Classification}}{Granularity of Origin Classification}}\label{granularity-of-origin-classification}

Our current prototype classifies execution origin into a small, fixed
set of categories (physical, remote, service, control-plane). While this
coarse granularity is intentional and sufficient for expressing many
practical policies, it may not capture all nuanced trust distinctions in
complex environments.

For example, remote graphical sessions, nested virtualization, or highly
customized access pathways may require additional refinement or explicit
marking. We view this not as a fundamental limitation, but as a policy
and deployment consideration.

\subsection{\texorpdfstring{\textbf{Dependence on Correct Control-Plane
Signaling}}{Dependence on Correct Control-Plane Signaling}}\label{dependence-on-correct-control-plane-signaling}

Distinguishing privileged control-plane access from arbitrary remote
access requires cooperation from trusted system components, such as PAM
modules, system services, or provider-specific agents. If such
components are misconfigured or compromised, origin classification may
be degraded.

However, this requirement aligns with existing trust assumptions in
managed environments, where control-plane software is already part of
the trusted computing base.

\subsection{\texorpdfstring{\textbf{Scope of
Protection}}{Scope of Protection}}\label{scope-of-protection}

Origin-aware enforcement restricts access to protected subsystems based
on execution origin. It does not prevent all forms of malicious behavior
by remote attackers. For example, attacks that operate entirely within
unprotected subsystems or exploit application-level vulnerabilities
remain out of scope.

Our approach is best understood as a mechanism for \emph{attack surface
reduction}, not as a comprehensive intrusion prevention system.

\subsection{\texorpdfstring{\textbf{Hardware and Input Trust
Assumptions}}{Hardware and Input Trust Assumptions}}\label{hardware-and-input-trust-assumptions}

The system does not attempt to distinguish between different types of
physical input devices (e.g., built-in keyboards versus USB devices).
Instead, it relies on kernel-visible execution context as the trust
boundary. While this simplifies design and improves robustness, it
assumes that local execution contexts are more trustworthy than remote
ones, an assumption consistent with many existing security models.

\subsection{\texorpdfstring{\textbf{Portability Across Operating
Systems}}{Portability Across Operating Systems}}\label{portability-across-operating-systems}

Although the concept of origin-aware access control is OS-agnostic, our
implementation targets Linux using eBPF LSM. Porting the mechanism to
other operating systems would require integration with their respective
kernel authorization frameworks and process metadata models.

Nevertheless, we believe the underlying abstraction is general and
applicable beyond Linux.

\subsection{\texorpdfstring{\textbf{Future
Work}}{Future Work}}\label{future-work}

While our prototype demonstrates the feasibility and practical value of
origin-aware access control, several extensions remain open for future
exploration.

\textbf{Origin-aware mediation of kernel control interfaces.} Our
prototype already applies origin-aware mediation to a subset of kernel
control interfaces, namely BPF syscalls mediated via the \texttt{bpf}
LSM hook, \textbf{demonstrating the feasibility of control-plane
mediation}. A natural extension is to generalize this approach to
additional privileged system calls such as \texttt{perf\_event\_open}
and other control-plane entry points. Enforcing origin-based
restrictions at these interfaces would further constrain post-compromise
attack surfaces, particularly those used for stealthy kernel
introspection or manipulation.

\textbf{Network-layer origin propagation.} In distributed and
containerized environments, execution origin may span multiple hosts or
orchestration layers. Future work could explore propagating origin
metadata across network boundaries---e.g., from a cloud control plane to
managed nodes---allowing origin-aware enforcement to distinguish between
actions initiated via trusted management channels and those originating
from untrusted remote sessions.

\textbf{Integration with existing MAC frameworks.} Although our design
operates as an orthogonal governance layer, tighter integration with
existing MAC frameworks such as SELinux or AppArmor remains an open
direction. Exposing execution origin as a first-class security attribute
to these systems could enable hybrid policies that combine fine-grained
labeling with coarse-grained origin constraints, while preserving the
centralized reasoning benefits of origin-aware enforcement.

\textbf{Formalization of origin-aware policy semantics.} Our current
policy model intentionally remains simple and imperative. Future work
could define a formal policy language for origin-aware rules, enabling
static analysis, conflict detection, and verification of origin-based
security invariants. Such a language could also facilitate reasoning
about policy composition with existing access control mechanisms.

\textbf{Stronger origin signals through hardware support.} While our
current prototype derives execution origin from kernel-maintained
software signals such as terminal association and process ancestry,
future systems could benefit from \textbf{hardware-assisted support for
securing origin metadata}. Prior work on hardware-assisted protections
argues that security-critical metadata should be integrity-protected,
efficiently accessible, and tightly coupled to execution state in order
to resist tampering and reduce enforcement overhead \cite{Jin2025}.
Execution origin in OAMAC naturally fits this model.

Hardware support could strengthen origin classification in two
complementary ways. First, hardware mechanisms could protect origin
metadata itself---ensuring that once an origin is assigned at task
creation, it cannot be forged or modified by compromised software
components. Second, hardware-assisted signals could provide stronger
provenance inputs, such as trusted physical presence indicators, secure
console devices, or attested management channels for cloud control-plane
access. If exposed to the kernel through trusted interfaces, such
mechanisms would enable unfakeable origin attribution while remaining
fully compatible with the origin-aware enforcement model proposed in
this work.

\section{\texorpdfstring{\textbf{Conclusion}}{Conclusion}}\label{conclusion}

We presented an origin-aware mandatory access control mechanism that
elevates \textbf{execution origin} to a first-class security attribute
in the Linux kernel. By distinguishing between physical user presence,
remote access, and service execution, our approach enables centralized
governance over security-critical subsystems and reduces post-compromise
attack surfaces that are difficult to control using identity- or
label-based mechanisms alone.

Our design is intentionally orthogonal to existing mandatory access
control frameworks. Rather than replacing systems such as SELinux or
AppArmor, origin-aware enforcement complements them by constraining
\emph{where} privileged actions may originate, independent of
application-specific policy. This separation allows administrators to
express high-level security intent without requiring deep subsystem
knowledge or extensive policy engineering.

We demonstrated that origin-aware access control can be implemented
today using upstream Linux features through the eBPF LSM framework,
requiring no kernel modifications and introducing minimal overhead.
Beyond enforcement, the origin layer provides a single, monitorable
control point whose integrity strongly correlates with the safety of the
protected subsystems, offering both practical protection and operational
visibility.

Taken together, our results suggest that execution origin represents a
missing abstraction in contemporary operating system security models.
Elevating this dimension enables practical attack surface reduction and
opens new directions for combining provenance-aware enforcement with
existing access control mechanisms in future systems.

\bibliographystyle{plainnat}
\bibliography{references}

\begin{thebibliography}{26}
\providecommand{\natexlab}[1]{#1}
\providecommand{\url}[1]{\texttt{#1}}
\expandafter\ifx\csname urlstyle\endcsname\relax
  \providecommand{\doi}[1]{doi: #1}\else
  \providecommand{\doi}{doi: \begingroup \urlstyle{rm}\Url}\fi

\bibitem[Ailabouni et~al.(2026)Ailabouni, Rom{\'a}n-Gallego, and
  P{\'e}rez-Delgado]{fg_rca}
Fouad Ailabouni, Jes{\'u}s-{\'A}ngel Rom{\'a}n-Gallego, and Mar{\'i}a-Luisa
  P{\'e}rez-Delgado.
\newblock Fg-rca: Kernel-anchored post-exploitation containment for iot with
  policy synthesis and mitigation of zero-day attacks.
\newblock \emph{IoT}, 7\penalty0 (1), 2026.
\newblock ISSN 2624-831X.
\newblock \doi{10.3390/iot7010003}.
\newblock URL \url{https://www.mdpi.com/2624-831X/7/1/3}.

\bibitem[{AppArmor Project}(2018)]{apparmor}
{AppArmor Project}.
\newblock Apparmor security project.
\newblock https://www.kernel.org/doc/Documentation/security/apparmor.txt, 2018.

\bibitem[B{\'e}lair et~al.(2021)B{\'e}lair, Laniepce, and Menaud]{snappy}
Maxime B{\'e}lair, Sylvie Laniepce, and Jean-Marc Menaud.
\newblock Snappy: Programmable kernel-level policies for containers.
\newblock In \emph{Proceedings of the 36th Annual ACM Symposium on Applied
  Computing}, SAC '21, pages 1636--1645, New York, NY, USA, 2021. Association
  for Computing Machinery.
\newblock ISBN 9781450381048.
\newblock \doi{10.1145/3412841.3442037}.
\newblock URL \url{https://doi.org/10.1145/3412841.3442037}.

\bibitem[Brimhall et~al.(2023)Brimhall, Garrard, De~La~Garza, and
  Coffman]{mac_survey}
Brennon Brimhall, Justin Garrard, Christopher De~La~Garza, and Joel Coffman.
\newblock A comparative analysis of linux mandatory access control policy
  enforcement mechanisms.
\newblock In \emph{Proceedings of the 16th European Workshop on System
  Security}, EUROSEC '23, pages 1--7, New York, NY, USA, 2023. Association for
  Computing Machinery.
\newblock ISBN 9798400700859.
\newblock \doi{10.1145/3578357.3589454}.
\newblock URL \url{https://doi.org/10.1145/3578357.3589454}.

\bibitem[{Cilium Project}(n.d.{\natexlab{a}})]{cilium}
{Cilium Project}.
\newblock Cilium: ebpf-based networking and security.
\newblock https://cilium.io, n.d.{\natexlab{a}}.

\bibitem[{Cilium Project}(n.d.{\natexlab{b}})]{tetragon}
{Cilium Project}.
\newblock Tetragon: Runtime security with ebpf.
\newblock GitHub repository, n.d.{\natexlab{b}}.
\newblock URL \url{https://github.com/cilium/tetragon}.
\newblock Accessed: 2026-01-20.

\bibitem[Costan and Devadas(2016)]{sgx}
Victor Costan and Srinivas Devadas.
\newblock Intel sgx explained.
\newblock Technical report, Cryptology ePrint Archive, Report 2016/086, 2016.
\newblock URL \url{https://eprint.iacr.org/2016/086.pdf}.

\bibitem[Costanzo(2025)]{ebpf_thesis}
Vincenzo Costanzo.
\newblock \emph{Detection and Mitigation of eBPF Security Risks in the Linux
  Kernel}.
\newblock PhD thesis, Politecnico di Torino, 2025.

\bibitem[Fournier et~al.(2021)Fournier, Afchain, and Baubeau]{ebpf_security}
Guillaume Fournier, Sylvain Afchain, and Sylvain Baubeau.
\newblock Runtime security monitoring with ebpf.
\newblock In \emph{SSTIC 2021}, 2021.
\newblock URL
  \url{https://www.sstic.org/media/SSTIC2021/SSTIC-actes/runtime_security_with_ebpf/SSTIC2021-Article-runtime_security_with_ebpf-fournier_afchain_baubeau.pdf}.

\bibitem[He et~al.(2023)]{bpf_attacks}
Zhibin He et~al.
\newblock Cross-container attacks: The bewildered ebpf on clouds.
\newblock In \emph{Proceedings of the 32nd USENIX Security Symposium}, 2023.
\newblock URL
  \url{https://www.usenix.org/system/files/usenixsecurity23-he.pdf}.

\bibitem[Jin et~al.(2025)Jin, Huang, Zeng, Xu, Wang, Lu, and Chen]{Jin2025}
Yue Jin, Tian-Yi Huang, Si-Yuan Zeng, Yi-Bin Xu, Han Wang, Tian-Yue Lu, and
  Ming-Yu Chen.
\newblock A survey of hardware-assisted intra-address space protections.
\newblock \emph{Journal of Computer Science and Technology}, 40\penalty0
  (5):\penalty0 1347--1367, September 2025.
\newblock ISSN 1860-4749.
\newblock \doi{10.1007/s11390-025-4827-z}.
\newblock URL \url{https://doi.org/10.1007/s11390-025-4827-z}.

\bibitem[Kerrisk and contributors(n.d.{\natexlab{a}})]{linux_capabilities}
Michael Kerrisk and contributors.
\newblock capabilities(7) — linux capabilities.
\newblock man7.org manual page, n.d.{\natexlab{a}}.
\newblock URL \url{https://man7.org/linux/man-pages/man7/capabilities.7.html}.
\newblock Accessed: 2026-01-20.

\bibitem[Kerrisk and contributors(n.d.{\natexlab{b}})]{namespaces}
Michael Kerrisk and contributors.
\newblock namespaces(7) — overview of linux namespaces.
\newblock man7.org manual page, n.d.{\natexlab{b}}.
\newblock URL \url{https://man7.org/linux/man-pages/man7/namespaces.7.html}.
\newblock Accessed: 2026-01-20.

\bibitem[Lim et~al.(2024)Lim, Prasad, Han, and Pasquier]{safebpf}
Soo~Yee Lim, Tanya Prasad, Xueyuan Han, and Thomas Pasquier.
\newblock Safebpf: Hardware-assisted defense-in-depth for ebpf kernel
  extensions.
\newblock In \emph{Proceedings of the 2024 on Cloud Computing Security
  Workshop}, CCSW '24, pages 80--94, New York, NY, USA, 2024. Association for
  Computing Machinery.
\newblock ISBN 9798400712340.
\newblock \doi{10.1145/3689938.3694781}.
\newblock URL \url{https://doi.org/10.1145/3689938.3694781}.

\bibitem[{LOLBAS Project}(n.d.)]{lolbins}
{LOLBAS Project}.
\newblock Lolbas: Living off the land binaries, scripts, and libraries, n.d.
\newblock URL \url{https://lolbas-project.github.io/}.
\newblock Accessed: 2026-01-20.

\bibitem[Loscocco and Smalley(2001)]{selinux}
Peter Loscocco and Stephen Smalley.
\newblock Security-enhanced linux.
\newblock NSA Technical Report, 2001.

\bibitem[Mazzocca et~al.(2025)Mazzocca, Garbugli, Armillotta, Montanari, and
  Bellavista]{ebpf_confinement}
Carlo Mazzocca, Andrea Garbugli, Michele Armillotta, Rebecca Montanari, and
  Paolo Bellavista.
\newblock Flexible and secure process confinement with ebpf.
\newblock In Fabio Martinelli and Ruben Rios, editors, \emph{Security and Trust
  Management}, pages 97--109. Springer Nature Switzerland, Cham, 2025.
\newblock ISBN 978-3-031-76371-7.

\bibitem[Miller et~al.(2019)Miller, Zhang, Zhuo, Xu, Krishnamurthy, and
  Anderson]{bpf_safety}
Samantha Miller, Kaiyuan Zhang, Danyang Zhuo, Shibin Xu, Arvind Krishnamurthy,
  and Thomas~E. Anderson.
\newblock Practical safe linux kernel extensibility.
\newblock In \emph{Proceedings of the 17th Workshop on Hot Topics in Operating
  Systems (HotOS '19)}. Association for Computing Machinery, 2019.
\newblock \doi{10.1145/3317550.3321429}.
\newblock URL \url{https://dl.acm.org/doi/10.1145/3317550.3321429}.

\bibitem[Mohammed(2026)]{oamac2026}
Omer Abdelmajeed~Idris Mohammed.
\newblock Oamac: Origin-aware mandatory access control for linux.
\newblock \url{https://github.com/omeroooor/oamac}, 2026.

\bibitem[Quinci et~al.(2025)Quinci, Belocchi, Quaglia, and Bianchi]{xfsm}
Arianna Quinci, Giacomo Belocchi, Francesco Quaglia, and Giuseppe Bianchi.
\newblock Stateful handling of critical events: Leveraging ebpf to realize
  extended finite state machine abstractions, 2025.

\bibitem[Schaufler(2008)]{smack}
Casey Schaufler.
\newblock Smack in embedded computing.
\newblock In \emph{Proceedings of the Ottawa Linux Symposium (OLS)}, pages
  179--186, 2008.
\newblock URL
  \url{https://landley.net/kdocs/ols/2008/ols2008v2-pages-179-186.pdf}.

\bibitem[Starovoitov(2015)]{ebpf_design}
Alexei Starovoitov.
\newblock Bpf in llvm and kernel.
\newblock In \emph{Linux Plumbers Conference}, 2015.
\newblock URL
  \url{https://blog.linuxplumbersconf.org/2015/ocw/sessions/3249.html}.

\bibitem[{Trusted Computing Group}(2019)]{tpm}
{Trusted Computing Group}.
\newblock Tpm 2.0 library specification, part 1: Architecture.
\newblock Specification, 2019.
\newblock URL
  \url{https://trustedcomputinggroup.org/resource/tpm-library-specification/}.
\newblock Accessed: 2026-01-20.

\bibitem[Wang et~al.(2024)Wang, Chen, Dai, Chen, Wei, and Zeng]{o2c}
Zicheng Wang, Tiejin Chen, Qinrun Dai, Yueqi Chen, Hua Wei, and Qingkai Zeng.
\newblock When ebpf meets machine learning: On-the-fly os kernel
  compartmentalization, 2024.
\newblock URL \url{https://arxiv.org/abs/2401.05641}.

\bibitem[Wright et~al.(2002)Wright, Cowan, Smalley, Morris, and
  Kroah-Hartman]{lsm_overview}
Chris Wright, Crispin Cowan, Stephen Smalley, James Morris, and Greg
  Kroah-Hartman.
\newblock Linux security modules: General security support for the linux
  kernel.
\newblock In \emph{11th USENIX security symposium (USENIX Security 02)}, 2002.

\bibitem[Yang et~al.(2024)Yang, Kang, and Nam]{Optimus2024}
S.~Yang, B.~B. Kang, and J.~Nam.
\newblock Optimus: association-based dynamic system call filtering for
  container attack surface reduction.
\newblock \emph{Journal of Cloud Computing}, 13:\penalty0 71, 2024.
\newblock \doi{10.1186/s13677-024-00639-3}.
\newblock URL \url{https://doi.org/10.1186/s13677-024-00639-3}.

\end{thebibliography}

\typeout{get arXiv to do 4 passes: Label(s) may have changed. Rerun}

\end{document}